\documentclass[lettersize,journal]{IEEEtran}
\usepackage{amsmath,amsfonts}
\usepackage{algorithmic}
\usepackage{algorithm}
\usepackage{array}
\usepackage[caption=false,font=normalsize,labelfont=sf,textfont=sf]{subfig}
\usepackage{textcomp}
\usepackage{stfloats}
\usepackage{url}
\usepackage{verbatim}
\usepackage{graphicx}
\usepackage{cite}
\usepackage{booktabs}
\usepackage{multicol}
\usepackage{graphicx}
\usepackage[export]{adjustbox}
\usepackage{array}
\usepackage{subcaption}
\usepackage{xcolor}
\newcolumntype{P}[1]{>{\centering\arraybackslash}p{#1}}

\usepackage{helvet} 

\usepackage{setspace} 

\usepackage{tikz}
\usetikzlibrary{shapes,arrows,positioning,calc}

\usepackage{afterpage}

\tikzset{
  block/.style = {rectangle, draw, fill=white, text centered, rounded corners, minimum height=2em, minimum width=1em}, 
  block2/.style = {rectangle, draw, fill=white, text centered, rounded corners, minimum height=2em, minimum width=1em},
  input/.style = {coordinate}, 
  output/.style = {coordinate},
  sum/.style = {draw, circle, inner sep=0pt, minimum size=0.5cm}, 
  sum2/.style = {draw, circle, inner sep=0pt, minimum size=0.5cm},
  arrow/.style = {->, thick, >=latex} 
}

\usepackage{hyperref}

\usepackage{bm}

\begin{document}

\bstctlcite{IEEEexample:BSTcontrol} 

\title{Intrinsic MIMO Particle Communication Channel with Random Advection}

\author{Fatih~Merdan,
       Ozgur B.~Akan,~\IEEEmembership{Fellow,~IEEE}

\thanks{Fatih Merdan and O.B. Akan are with the Center for neXt-generation Communications (CXC), Department of of Electrical and Electronics Engineering, Koç University, Istanbul 34450, Türkiye (e-mail:  fmerdan25@ku.edu.tr, akan@ku.edu.tr).}
\thanks{O. B. Akan is also with the Internet of Everything (IoE) Group,
Department of Engineering, University of Cambridge, Cambridge CB3 0FA,
U.K. (e-mail: oba21@cam.ac.uk).}}

\markboth{}%
{}

\maketitle

\begin{abstract}
In this work, receiver diversity in advection-dominated diffusion–advection channels is investigated. Strong directed flow fundamentally alters the communication-theoretic properties of molecular communication systems (MC). Specifically, advection preserves the temporal ordering and shape of transmitted pulses, enabling pulse-based and higher-order modulation schemes that are typically infeasible in purely diffusive environments. Focusing on a single transmitter and a single type of information molecule, it is demonstrated that spatially distributed receivers can observe distinct realizations of the same transmitted signal, giving rise to diversity gain. Several receiver combining strategies are evaluated and shown to improve detection performance compared to single-receiver operation, particularly in low-to-moderate signal-to-noise ratio (SNR) regimes. The results provide a structured framework for understanding receiver-side diversity in molecular communication, highlighting the role of advection as a key enabler for reliable pulse-based signaling. This perspective establishes a foundation for future studies on advanced modulation, joint equalization and detection, and multi-molecule MIMO extensions that can further enhance the performance and physical applicability of MC systems. 

\end{abstract}


\section{Introduction}
\IEEEPARstart{M}{olecular} communication (MC) has emerged as a promising paradigm for information exchange in environments where conventional electromagnetic communication is infeasible, such as biological systems and microfluidic platforms \cite{A_Comprehensive_Survey_of_Recent_Advancements_in_Molecular_Communication}. In MC, information is encoded into the release patterns of particles or molecules, which propagate through a medium via physical transport mechanisms and are detected by receivers through chemical sensing processes. Unlike conventional communication systems, MC is fundamentally governed by diffusion, advection, and reaction dynamics, leading to unique challenges and opportunities in system design and performance analysis \cite{Modeling_ConvectionDiffusionReaction_Systems_for_Microfluidic_Molecular_Communications_with_SurfaceBased_Receivers_in_Internet_of_BioNano_Things}.

Early studies in MC predominantly focused on diffusion-driven channels, where molecular propagation is isotropic and inherently slow. In such settings, severe intersymbol interference (ISI), long channel memory, and limited controllability of signal shape significantly restrict achievable data rates and detection accuracy. As a result, much of the existing literature has concentrated on mitigating ISI through coding, detection, and pulse-shaping techniques, as well as on developing analytical channel models under purely diffusive assumptions \cite{Maximum_Likelihood_Detection_With_Ligand_Receptors_for_DiffusionBased_Molecular_Communications_in_Internet_of_BioNano_Things}. In many practical and biological environments, however, molecular transport is not purely diffusive. Fluid flow induced by pressure gradients, convection, or environmental motion introduces advection. Until now, advection has been mostly discussed as a means of increasing the speed of propagation and reducing the channel memory \cite{A_Molecular_Communication_Platform_Based_on_Body_Area_Nanonetwork}, and as a constant property of the channel \cite{Diffusive_Molecular_Communication_with_Disruptive_Flows}, \cite{Channel_Modeling_for_Diffusive_Molecular_CommunicationA_Tutorial_Review}. However, random advection with a strong mean direction can preserve pulse ordering, making advection-dominated MC particularly attractive for pulse-based modulation schemes, where information is conveyed through the timing, shape, or orthogonality of released molecular pulses \cite{Airborne_Particle_Communication_Through_Timevarying_DiffusionAdvection_Channels}.

A MIMO communication link in MC was first demonstrated in \cite{Molecular_MIMO_communication_link}. Later, a similar experiment with a more detailed analysis was provided in \cite{Molecular_MIMO_From_Theory_to_Prototype}. The following studies on this topic focused on advanced modulation and coding techniques for diffusive topologies \cite{Spatial_Modulation_for_Molecular_Communication}, \cite{Molecular_Type_Permutation_Shift_Keying_in_Molecular_MIMO_Communications_for_IoBNT}. Although there exist some studies that consider MIMO MC in the presence of drift \cite{Evolutionary_generative_adversarial_network_based_end_to_end_learning_for_MIMO_molecular_communication_with_drift_system}, these approaches are largely data-driven or experimental in nature and do not provide a systematic analytical characterization. In particular, the implications of advection-dominated transport on the fundamental structure of MIMO MC channels remain largely unexplored. This paper addresses this gap by examining advection-dominated diffusion–advection channels, where a strong, directed flow preserves the temporal ordering and shape of transmitted pulses. This property enables the use of pulse-based and higher-order modulation schemes that are generally impractical in purely diffusive MC environments. It is demonstrated that under directed advection, the spatial dispersion induced by random transverse flow results in multiple distinct receiver observations of the same transmitted pulse. This perspective motivates an interpretation of MC as an intrinsically MIMO system, where diversity arises naturally from spatial sampling of the particle cloud rather than from multiple transmitters or molecule types. 

The purpose of this study is to provide a structured framework for understanding how receiver diversity can be exploited in pulse-based molecular communication under random advection. Several receiver combining strategies are evaluated, and their impact on bit error rate (BER) is studied through Monte Carlo simulations. Results demonstrate that multi-receiver diversity can improve detection performance, especially in low-to-moderate signal-to-noise ratio (SNR) regimes, even when individual receiver observations are highly noisy.

The remainder of the paper is organized as follows. Section II discusses the intrinsic MIMO interpretation of molecular communication and clarifies the distinction between diversity and multiplexing gains in MC. Section III introduces the system model, including the advection–diffusion channel and pulse-based signaling framework. Section IV presents simulation results and performance comparisons for different combining schemes. Finally, Section V concludes the paper and outlines directions for future research.

\section{Intrinsic MIMO Property of Molecular Communication}

In conventional communication, MIMO offers two types of gain: diversity gain, where accuracy improves, and multiplexing gain, where the rate of communication improves.
In MC, multiple data streams cannot be sent simultaneously using only one type of molecule, since concentrations due to different data streams would mix with each other. This may be possible if time division multiplexing (TDM) is at play, but this is equivalent to just combining the data streams into one data stream in a structured manner. Trying to use the same particle to send two different data streams is like using the same frequency interval to send data simultaneously. In MC, there is no resource like a frequency spectrum. The closest thing is the use of different types of molecules, which can be seen as using different frequency intervals. Obtaining independent frequency channels would correspond to using molecules that do not chemically interact with each other and that do not affect the movements of each other under diffusion-advection channels. However, having multiple types of information molecules does not necessarily mean multiplexing. For example, it is proposed to use ratios of concentrations of different molecules to send a message \cite{Ratio_Shift_Keying_Modulation_for_TimeVarying_Molecular_Communication_Channels}. This falls into the concept of pulse shape design, not MIMO. Therefore, there are different ways to utilize multiple particle types as well. In any case, to properly understand the multiplexing gain in MC, one has to consider different types of molecules. This article focuses only on the diversity gain in MC, and only a single type of molecule is used.

\begin{figure*}[!t]
  \centering
  \includegraphics[width=0.9\linewidth]{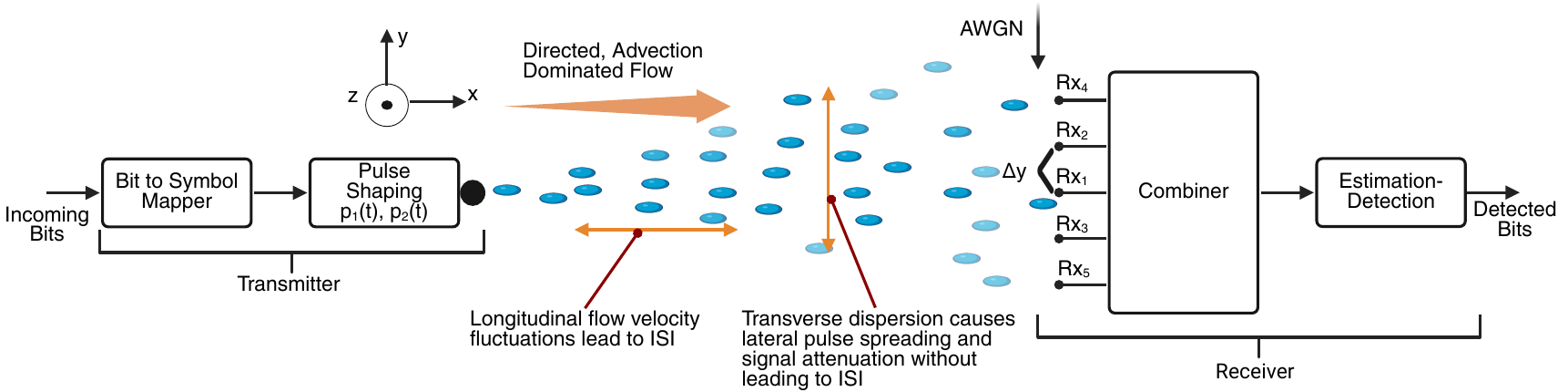}
  \caption{Block diagram of the proposed end-to-end molecular communication framework under directed advection, illustrating symbol generation and pulse emission at the transmitter, particle transport, and multi-receiver combining and detection at the receiver side. \cite{BioRender_Merdan_2026_2}.}
  \label{Part_Comm_Diversity_System_Diagram}
\end{figure*}

With only one type of molecule, having multiple transmitters is functionally equivalent to having one transmitter whose output is an amplified version of the individual transmitters. This is true assuming that the information molecule is released to the channel from a fixed position, so that each transmitter just scales up the signal. In contrast, spatially distributed transmitters can introduce geometric and flow-induced diversity, opening the door to new transmission design strategies. However, a rigorous analysis of such networks requires higher-layer and network-level modeling, which is beyond the scope of this paper. This is why only a single transmitter is used in this article.

The previous discussion indicates that understanding MIMO in MC starts with understanding how multiple receivers might be useful for communication. Even in the case where only one transmitter is used, the information is spread across all space through diffusion and advection. This points out that the molecules released from one transmitter can correspond to many different channels if there are properly located receivers. Moreover, the fact that the signals are always nonnegative and there are no phase-coherence technicalities in MC suggests that each small channel can be useful. This is the main difference between the unconventional and conventional communication topologies. The notion that every molecule can correspond to a single different channel when there is a receiver to read it naturally leads to the interpretation of molecular communication as an \textit{intrinsically MIMO} system. In this article, it is shown that with this intrinsic MIMO, a diversity gain can be obtained.

\section{System Model}

Complex modulation schemes can be implemented using orthogonal pulses in flows dominated by advection to transform the incoming bit sequence into a symbol sequence \cite{Airborne_Particle_Communication_Through_Timevarying_DiffusionAdvection_Channels}. For this purpose, the symbol duration is partitioned into $N$ non-overlapping subintervals, each associated with one rectangular pulse that is active only within its assigned subinterval. These pulses are mutually orthogonal in time and span an $N$-dimensional signal space. Information is conveyed by selecting one of $M$ discrete amplitude levels independently along each dimension, resulting in a constellation of $M^N$ possible symbols. With this setup, high cardinality modulations can be studied to observe the benefits of the multi-receiver system easily.

In \cite{Airborne_Particle_Communication_Through_Timevarying_DiffusionAdvection_Channels}, an analytical solution is provided for the channel impulse response for a particle-based communication under time-varying advection. The same model is used in this article, where the flow of advection is assumed to be the same at all positions. Furthermore, advection velocities through dimensions $x$ and $y$  are modeled as independent Gaussian random variables, and zero advection is assumed for dimension $z$. There is a directed flow in the $+x$ direction, and multiple receivers are placed at the same $x$ position with $\Delta y$ separation in their $y$ positions as shown in \autoref{Part_Comm_Diversity_System_Diagram}. 

Using orthogonal pulses is possible in this topology because their order and shapes are preserved when they reach the receiver due to strong advection. As summarized in \autoref{Part_Comm_Diversity_System_Diagram}, the $x$ and $y$ components of the flow velocity have different effects on the performance of the given communication system. The stochastic behaviour of the $x$ component of the flow velocity can cause consecutive pulses to mix with each other, resulting in severe intersymbol interference (ISI). On the other hand, random characteristics of the $y$ component of the flow velocity reduce the signal power a receiver can get without causing any ISI. However, it can still change the shape of the pulses and distort the structure of the constellation, causing errors. This observation highlights that if one uses only one receiver, there can be many more molecules carrying the same message around this receiver. This behavior motivates the use of multiple spatially distributed receivers. With multiple receivers, the system can read the same message through other sensors. If one can combine all these readings of the same message in an intelligent manner, she can improve the performance of the communication system. This is directly analogous to receiver diversity combining in classical wireless communication systems. 

$Rx_1$ in \autoref{Part_Comm_Diversity_System_Diagram} is labeled the main receiver since the mean flow carries molecules to this receiver. The receivers are assumed to be fully time-synchronized, and the clock of the main receiver is used. Each receiver works as explained in \cite{Airborne_Particle_Communication_Through_Timevarying_DiffusionAdvection_Channels}. After each receiver applies sampling with the symbol period, the results are sent to the combiner, where diversity combining techniques are applied. The output of the combiner goes to the estimation-detection block. Here, receivers use automatic gain control (AGC) for correct scaling, and then decisions are made. The diffusive channel receiver design was previously discussed in \cite{Receiver_Design_for_Molecular_Communication}. However, to the best of our knowledge, currently, there is no such study on time-varying diffusion-advection channels. Therefore, there may exist better ways to do estimation and detection than the ones adopted in this article. Addressing these open problems is essential for realizing the practical implementation of MC systems under directed advection.

Similar to conventional wireless systems, different diversity combining techniques can be implemented. Four different such techniques are implemented in this article. These are Selection Combining (SC), Equal Gain Combining (EGC), Distribution Gain Combining (DGC), and Pilot Energy Gain Combining (PGC).

\textbf{Selection Combining (SC):} Corresponds to directly choosing the main receiver. This is not a diversity combining, and it is included for comparison purposes.

\textbf{Equal Gain Combining (EGC):} Each receiver output is added together using the same normalized weights.

\textbf{Distribution Gain Combining (DGC):} Using the $y$ coordinate of each receiver, The distribution value of the $y$ component of the flow velocity is calculated. These are normalized and used as weights in the weighted sum, which constitutes the output of the combiner.

\textbf{Pilot Energy Gain Combining (PGC):} Pilot energies for each receiver are calculated and normalized and used as weights in the weighted sum.

To obtain a diversity gain, the side receivers must be sufficiently close to the main receiver such that they observe pulse-structured signals. If the transverse separation $\Delta y$ is chosen too large—particularly when the variance of the transverse flow velocity component is small—some receivers may fail to observe any meaningful signal, or may capture only a truncated portion of the transmitted pulse. In such cases, EGC aggregates additional noise without contributing structured signal energy, and therefore, no diversity gain is expected. The notion of a structured signal is treated as a system design parameter. A practical and implementation-oriented definition can be based on the received pilot symbol energies. For a side receiver $Rx_j$, the total pilot symbol energy is written as 
\begin{equation}
\label{total_pilot_energy}
\begin{split}
E_j = \sum_{k=1}^{N_{\mathrm{pilot}}} \bigl\lVert \mathbf{w}_{j,k} \bigr\rVert^2 ,
\end{split}
\end{equation}

\noindent where $\mathbf{w}_{j,k}$ denotes the matched-filter output vector corresponding to the $k^{th}$ pilot symbol at receiver $Rx_j$. For each side receiver, the pilot energy ratio is written as $\rho_j = \frac{E_j}{E_{\mathrm{main}}},$ where $E_{\mathrm{main}}$ denotes the total pilot energy at the main receiver. The observation at receiver $Rx_j$ is declared structured if $\rho_j \geq \eta$ for some design threshold $0 \leq \eta \leq 1$. Since $E_j$ (and hence $\rho_j$) is random due to the stochastic nature of the diffusion–advection channel, a critical transverse distance $y_{c}$ can be defined probabilistically as
\begin{equation}
\label{crit_distance}
\begin{split}
y_c = \sup \Bigl\{\, |y| \;:\; \mathbb{P}\!\left( \rho_j \ge \eta \right) \ge 1 - \delta \Bigr\},
\end{split}
\end{equation}

\noindent where $\delta \in (0,1)$ specifies the allowable probability of misclassification. Here, $y_{c}$ represents the maximum transverse separation from the main receiver for which a side receiver contributes reliably to diversity gain.

\section{Simulation Results}

The simulation parameters are given in \autoref{sim_params}. The simulations were performed in MATLAB R2023b.

\begin{table}[h]
\centering
\caption{Parameters Used in the Simulations}
\label{sim_params}
\setlength{\tabcolsep}{4pt}
\renewcommand{\arraystretch}{1.15}
\begin{tabular}{l c}
\hline
\textbf{Parameter} & \textbf{Value} \\
\hline
\multicolumn{2}{c}{\textit{Communication}} \\
\hline
Modulations & (N,M)=(2,4),(3,3),(3,4),(4,2) \\
Symbol duration $T_{\mathrm{sym}}$ & $2~\mathrm{s}$ \\
Pilots / data symbols & $N_{\mathrm{pilot}}=$ $32$ / $N_{\mathrm{data}}=$ $1000$ \\
\hline
\multicolumn{2}{c}{\textit{Sampling \& Channel Memory}} \\
\hline
Channel sampling $f_{\mathrm{sim}}$ & $1000~\mathrm{Hz}$ \\
RX processing rate $f_{\mathrm{rx}}$ & $100~\mathrm{Hz}$ \\
Channel memory $T_{\mathrm{mem}}$ & $30~\mathrm{s}$ \\
\hline
\multicolumn{2}{c}{\textit{Physical Channel}} \\
\hline
Diffusion coefficient $D$ & $6.7698 \times 10^{-6}~\mathrm{m^2/s}$ \\
Mean wind (x) $\mu_{v_x}$ & $0.5~\mathrm{m/s}$ \\
Std. wind (x) $\sigma_{v_x}$ & $10^{-3}~\mathrm{m/s}$ \\
Mean wind (y) $\mu_{v_y}$ & $0~\mathrm{m/s}$ \\
Std. wind (y) $\sigma_{v_y}$ & $0.1~\mathrm{m/s}$ \\
\hline
\multicolumn{2}{c}{\textit{Geometry}} \\
\hline
Transmitter position & $(0,\,0,\,1) \mathrm{m}$ \\
Main receiver position & $(1,\,0,\,1) \mathrm{m}$ \\
\hline

\end{tabular}
\end{table}

\begin{figure}[h]
  \centering
  \includegraphics[width=0.75\linewidth]{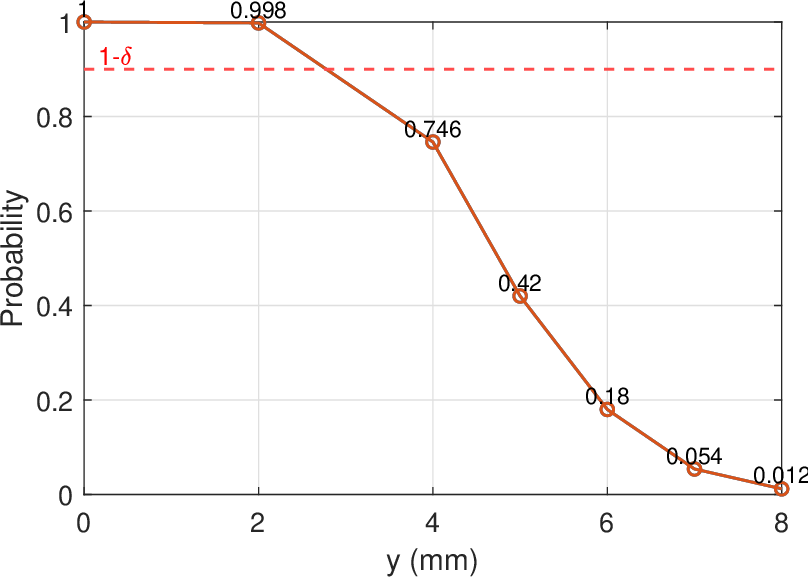}
  \caption{Monte-Carlo estimate of the structured-signal probability $\mathbb{P}(\rho_j \ge \eta)$ versus transverse receiver position $y$. The dashed line indicates the target probability level $1-\delta$.}
  \label{StructuredProb_vs_x2_eta_0p7_delta_0p1_Nmc_500}
\end{figure}

For receivers placed at different transverse positions, the probability $\mathbb{P}\!\left( \rho_j \ge \eta \right)$ is estimated via Monte Carlo simulations. To do this, the modulation scheme $(N,M)=(2,4)$ is employed at $SNR=-5dB$ with $\eta = 0.7$ and $\delta=0.1$. For each receiver location, the transmission is repeated 500 times to evaluate whether the condition $\mathbb{P}\!\left( \rho_j \ge \eta \right) \ge 1 - \delta$ is satisfied. Only positive $y$ are used since there is symmetry. The resulting probability estimates are given in \autoref{StructuredProb_vs_x2_eta_0p7_delta_0p1_Nmc_500}.

\begin{figure}[h]
  \centering

  \subfloat[]{
    \includegraphics[width=0.5\linewidth]{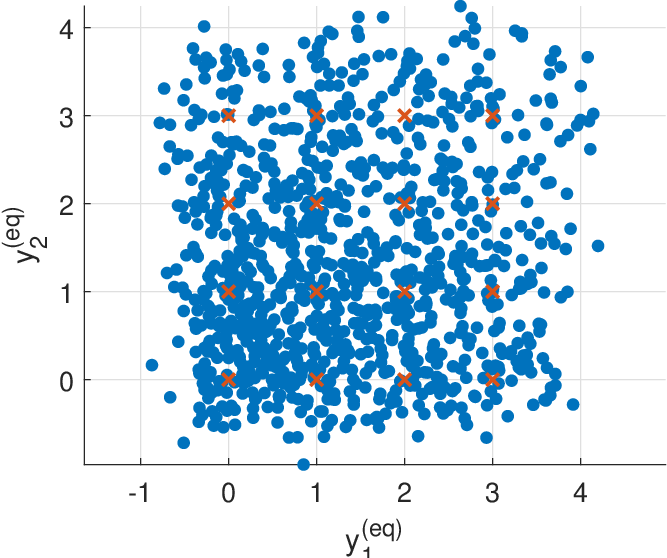}
    \label{Equalized_Overlay_MAIN_ONLY_Mode_SNR_In_m5_Mod_6_Tsym_2_FsSim_1000_FsTx_100_MuX_0p5_SigX_0p001_MuY_0_SigY_0p1_Nr_5}}
  \subfloat[]{
    \includegraphics[width=0.5\linewidth]{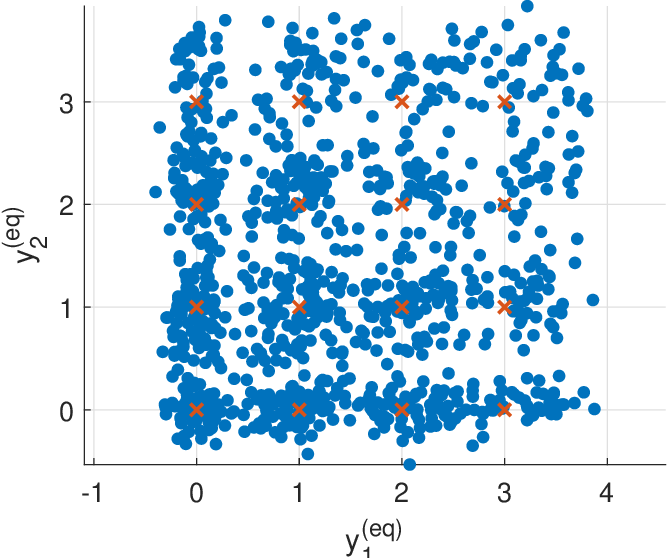}
    \label{Equalized_Overlay_EGC_ALL_Mode_SNR_In_m5_Mod_6_Tsym_2_FsSim_1000_FsTx_100_MuX_0p5_SigX_0p001_MuY_0_SigY_0p1_Nr_5}}
  \\[0.8em]
  \subfloat[]{
    \includegraphics[width=0.5\linewidth]{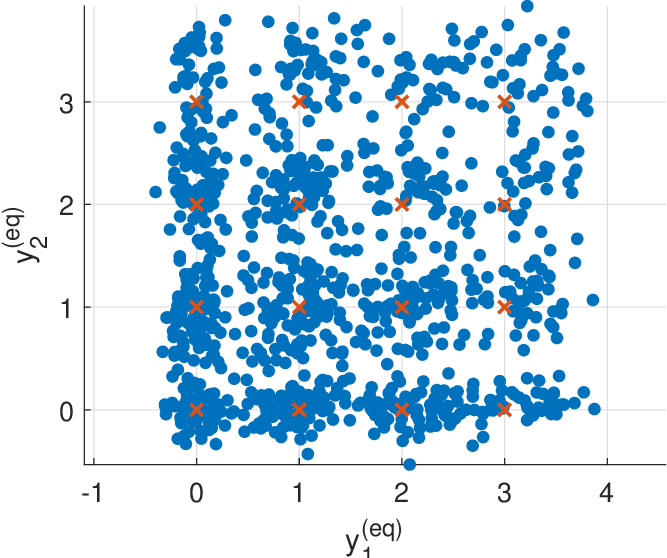}
    \label{Equalized_Overlay_DIST_GAUSS_WEIGHTED_Y_Mode_SNR_In_m5_Mod_6_Tsym_2_FsSim_1000_FsTx_100_MuX_0p5_SigX_0p001_MuY_0_SigY_0p1_Nr_5}}
  \subfloat[]{
    \includegraphics[width=0.5\linewidth]{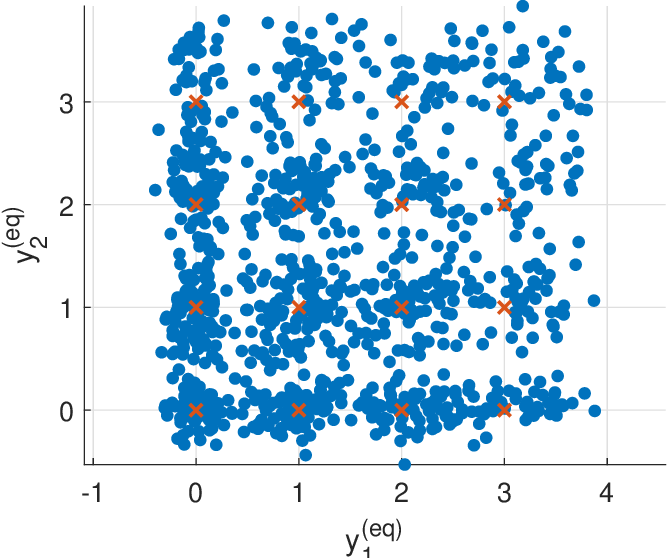}
    \label{Equalized_Overlay_PILOT_ENERGY_WEIGHTED_Mode_SNR_In_m5_Mod_6_Tsym_2_FsSim_1000_FsTx_100_MuX_0p5_SigX_0p001_MuY_0_SigY_0p1_Nr_5}}

  \caption{Constellation diagrams of the modulation scheme $(N,M)=(2,4)$ under -5 dB SNR for
  \protect\subref{Equalized_Overlay_MAIN_ONLY_Mode_SNR_In_m5_Mod_6_Tsym_2_FsSim_1000_FsTx_100_MuX_0p5_SigX_0p001_MuY_0_SigY_0p1_Nr_5} the main receiver only (BER = 0.149), 
  \protect\subref{Equalized_Overlay_EGC_ALL_Mode_SNR_In_m5_Mod_6_Tsym_2_FsSim_1000_FsTx_100_MuX_0p5_SigX_0p001_MuY_0_SigY_0p1_Nr_5} Equal Gain Combining (BER = 0.09956),
  \protect\subref{Equalized_Overlay_DIST_GAUSS_WEIGHTED_Y_Mode_SNR_In_m5_Mod_6_Tsym_2_FsSim_1000_FsTx_100_MuX_0p5_SigX_0p001_MuY_0_SigY_0p1_Nr_5} Distribution Gain Combining (BER = 0.09956) and
  \protect\subref{Equalized_Overlay_PILOT_ENERGY_WEIGHTED_Mode_SNR_In_m5_Mod_6_Tsym_2_FsSim_1000_FsTx_100_MuX_0p5_SigX_0p001_MuY_0_SigY_0p1_Nr_5} Pilot Energy Gain Combining (BER = 0.09811).}
  \label{Constellaiton_comparison}
\end{figure}

Using the result in \autoref{StructuredProb_vs_x2_eta_0p7_delta_0p1_Nmc_500}, $4$ side receivers are positioned at y = $\pm 0.001 \, m, \pm0.002 \, m$. For $SNR = -5 dB$, combining methods are applied and the resulting constellation diagrams of the modulation scheme $(N,M)=(2,4)$ are given in \autoref{Constellaiton_comparison}. It can be observed that the constellations obtained using (EGC), (DGC), and (PGC) are more compact, and the symbols are more distinguishable compared to the single-receiver case, and they perform better in terms of BER. In general, (EGC), (DGC), and (PGC) seem to perform similarly in this configuration.

\begin{figure}[h]
  \centering

  \subfloat[]{
    \includegraphics[width=0.5\linewidth]{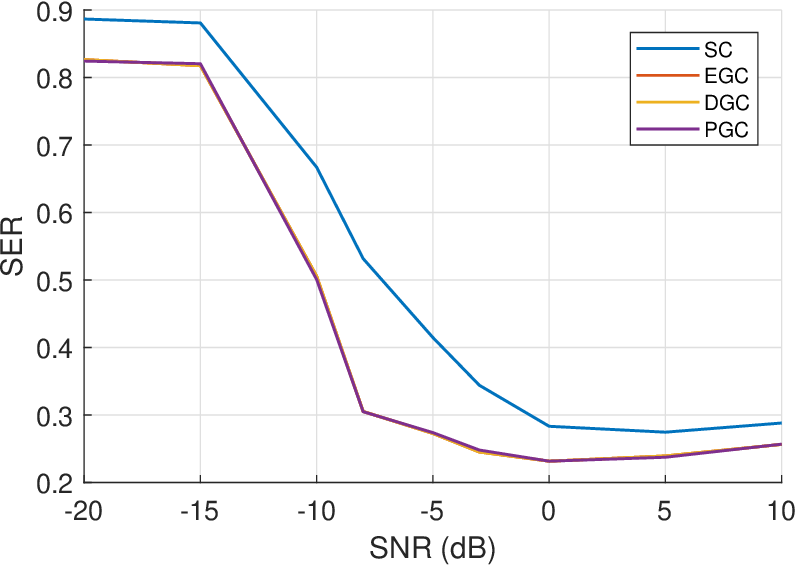}
    \label{SER_SNR_AVG_Ndim_2_M_4_Eq_mmse_Tsym_NaN_FsSim_NaN_FsTx_NaN_Nr_NaN}}
  \subfloat[]{
    \includegraphics[width=0.5\linewidth]{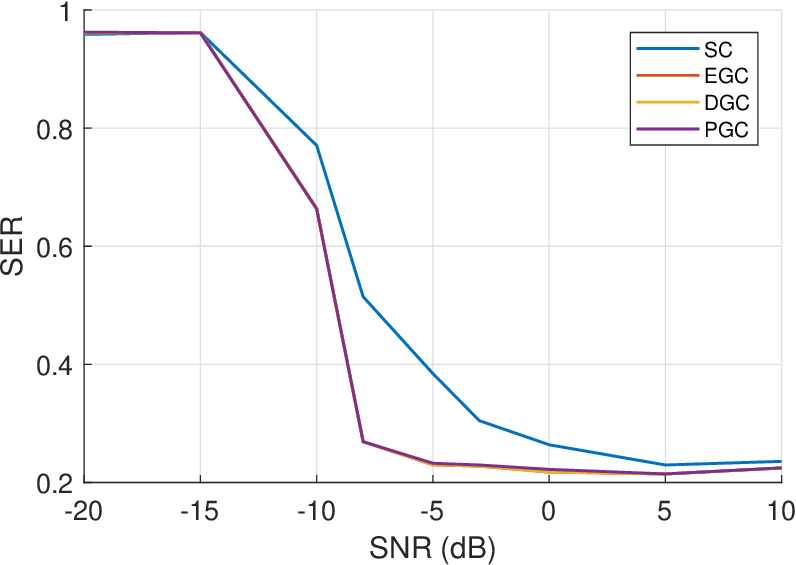}
    \label{SER_SNR_AVG_Ndim_3_M_3_Eq_mmse_Tsym_NaN_FsSim_NaN_FsTx_NaN_Nr_NaN}}
  \\[0.8em]
  \subfloat[]{
    \includegraphics[width=0.5\linewidth]{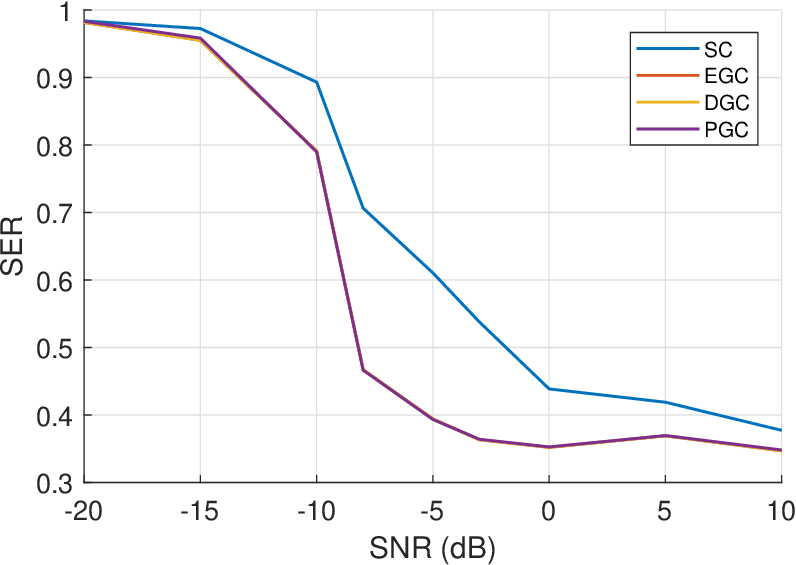}
    \label{SER_SNR_AVG_Ndim_3_M_4_Eq_mmse_Tsym_NaN_FsSim_NaN_FsTx_NaN_Nr_NaN}}
  \subfloat[]{
    \includegraphics[width=0.5\linewidth]{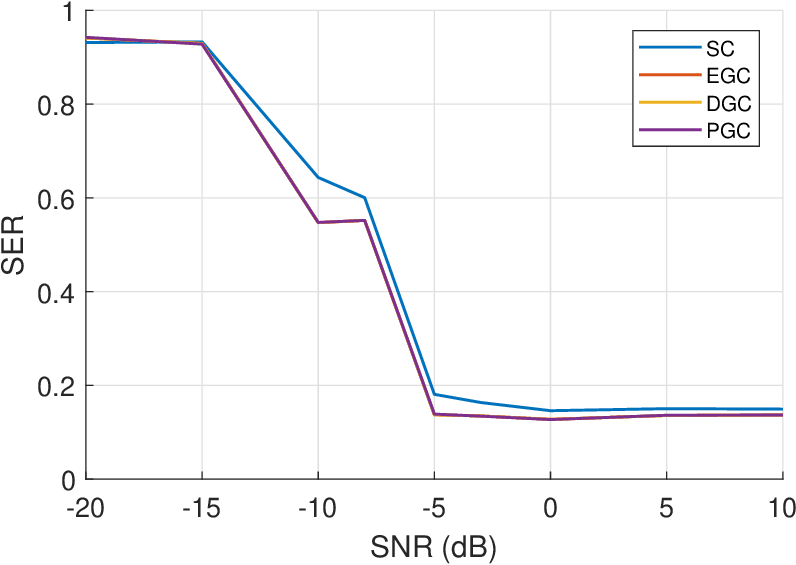}
    \label{SER_SNR_AVG_Ndim_4_M_2_Eq_mmse_Tsym_NaN_FsSim_NaN_FsTx_NaN_Nr_NaN}}

  \caption{SER performance versus SNR of the considered combining schemes for modulations
  \protect\subref{SER_SNR_AVG_Ndim_2_M_4_Eq_mmse_Tsym_NaN_FsSim_NaN_FsTx_NaN_Nr_NaN} (N,M)=(2,4), 
  \protect\subref{SER_SNR_AVG_Ndim_3_M_3_Eq_mmse_Tsym_NaN_FsSim_NaN_FsTx_NaN_Nr_NaN} (N,M)=(3,3),
  \protect\subref{SER_SNR_AVG_Ndim_3_M_4_Eq_mmse_Tsym_NaN_FsSim_NaN_FsTx_NaN_Nr_NaN} (N,M)=(3,4) and
  \protect\subref{SER_SNR_AVG_Ndim_4_M_2_Eq_mmse_Tsym_NaN_FsSim_NaN_FsTx_NaN_Nr_NaN} (N,M)=(4,2).}
  \label{Modulation_comparison}
\end{figure}

Using the same $5$ receivers and the MMSE equalization, diversity combining methods are applied  on different modulation schemes for different SNR regimes  and SER is calculated. The resulting plot is given in \autoref{Modulation_comparison}. In general, SC exhibits the highest error rates across the entire SNR range, indicating limited robustness against stochastic advection and noise. In contrast, multi-receiver combining significantly improves performance, particularly in the low-to-moderate SNR regime. The gains diminish at higher SNR values, where noise becomes less dominant, and all schemes converge to similar error levels. These results also indicate that diversity gain can be obtained for different modulation schemes.

\begin{figure}[h]
  \centering

  \subfloat[]{
    \includegraphics[width=0.5\linewidth]{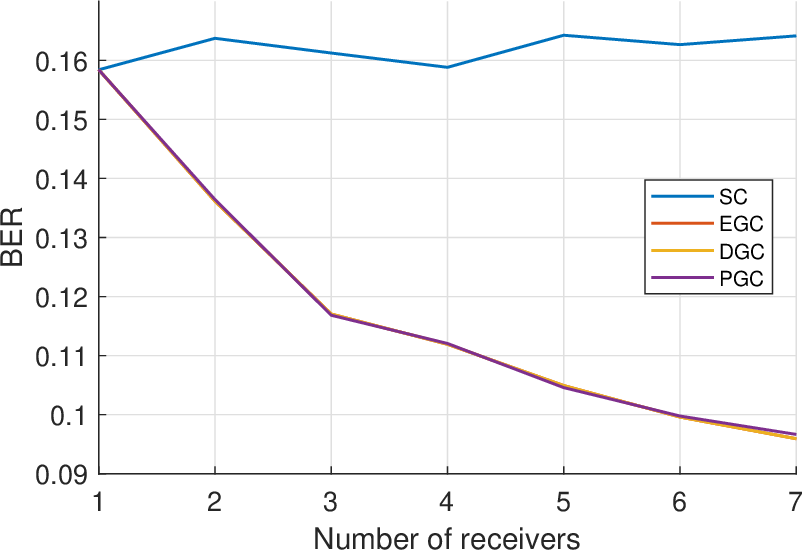}
    \label{BER_vs_Nr_AVG_SNR_m5_Mod_6_Tsym_2_Dy_0p001_Eq_structured}}
  \subfloat[]{
    \includegraphics[width=0.5\linewidth]{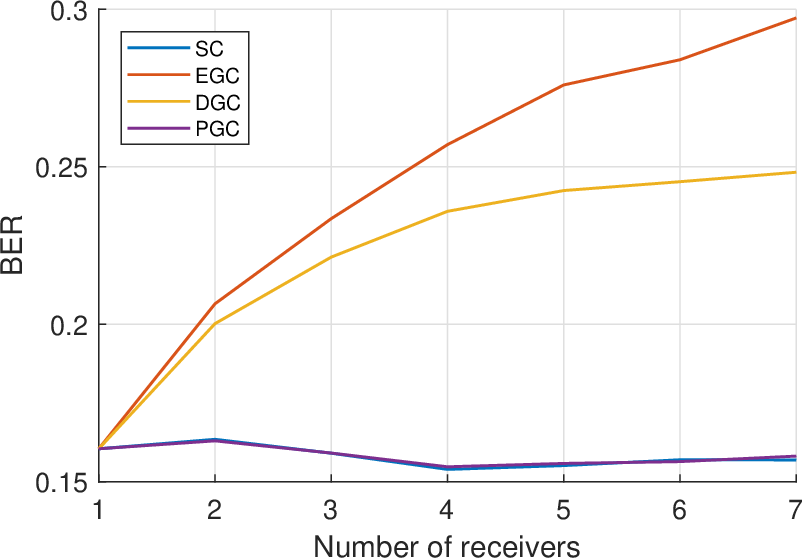}
    \label{BER_vs_Nr_AVG_SNR_m5_Mod_6_Tsym_2_Dy_0p05_Eq_NONstructured}}
    
  \caption{BER performance versus the number of receivers of the considered combining schemes for the modulation (N,M)=(2,4) where the channel output at the side receivers are
  \protect\subref{BER_vs_Nr_AVG_SNR_m5_Mod_6_Tsym_2_Dy_0p001_Eq_structured} structured ($\Delta y = 0.001\, m$)
  \protect\subref{BER_vs_Nr_AVG_SNR_m5_Mod_6_Tsym_2_Dy_0p05_Eq_NONstructured} non-structured ($\Delta y = 0.05 \, m$).}
  \label{BER_vs_NRx_structured_vs_nonstructured}
\end{figure}

To study the impact of spatial diversity as a function of the number of receivers ($N_r$), $N_r$ is varied at a fixed $SNR = -5dB$ while keeping the combining methods unchanged. Receivers are placed symmetrically along the transverse direction, and two cases are considered: (a) a \emph{structured} regime where all receivers observe structured signals, and (b) a \emph{non-structured} regime where some receivers observe severely attenuated or truncated signals. As shown in \autoref{BER_vs_NRx_structured_vs_nonstructured}\protect\subref{BER_vs_Nr_AVG_SNR_m5_Mod_6_Tsym_2_Dy_0p001_Eq_structured}, increasing $N_r$ in the structured regime yields a clear BER improvement, confirming the presence of diversity gain. In contrast, \autoref{BER_vs_NRx_structured_vs_nonstructured}\protect\subref{BER_vs_Nr_AVG_SNR_m5_Mod_6_Tsym_2_Dy_0p05_Eq_NONstructured} shows that in the non-structured regime, adding receivers can degrade performance, since unstructured observations contribute noise without useful signal energy. However, PGC can eliminate these receivers by assigning small weights to them.

\section{Conclusion}

This work examined receiver diversity in MC over advection-dominated diffusion–advection channels. Strong directed advection was shown to preserve the temporal ordering and shape of transmitted pulses, thereby enabling pulse-based and higher-order modulation schemes that are typically impractical in purely diffusive environments. By restricting attention to a single transmitter and a single molecule type, the analysis deliberately isolated the effect of spatial receiver diversity and showed that meaningful performance gains can be obtained without invoking multi-molecule signaling or distributed transmission. 

The framework developed in this study provides a foundation for several important extensions. Future work may investigate alternative modulation schemes tailored to advection-dominated channels, more advanced combining techniques in which equalization and detection are jointly optimized, and receiver-side algorithms that explicitly account for channel memory. In addition, extending the analysis to multiple molecule types introduces the possibility of true multiplexing gains, complementing the diversity gains studied here. Such directions are essential for bridging the gap between physically realizable molecular transport mechanisms and scalable, high-performance MC systems.

\bibliographystyle{IEEEtran}
\bibliography{references.bib}

\vspace{11pt}

\begin{IEEEbiography}[{\includegraphics[width=1in,height=1.25in,clip,keepaspectratio]{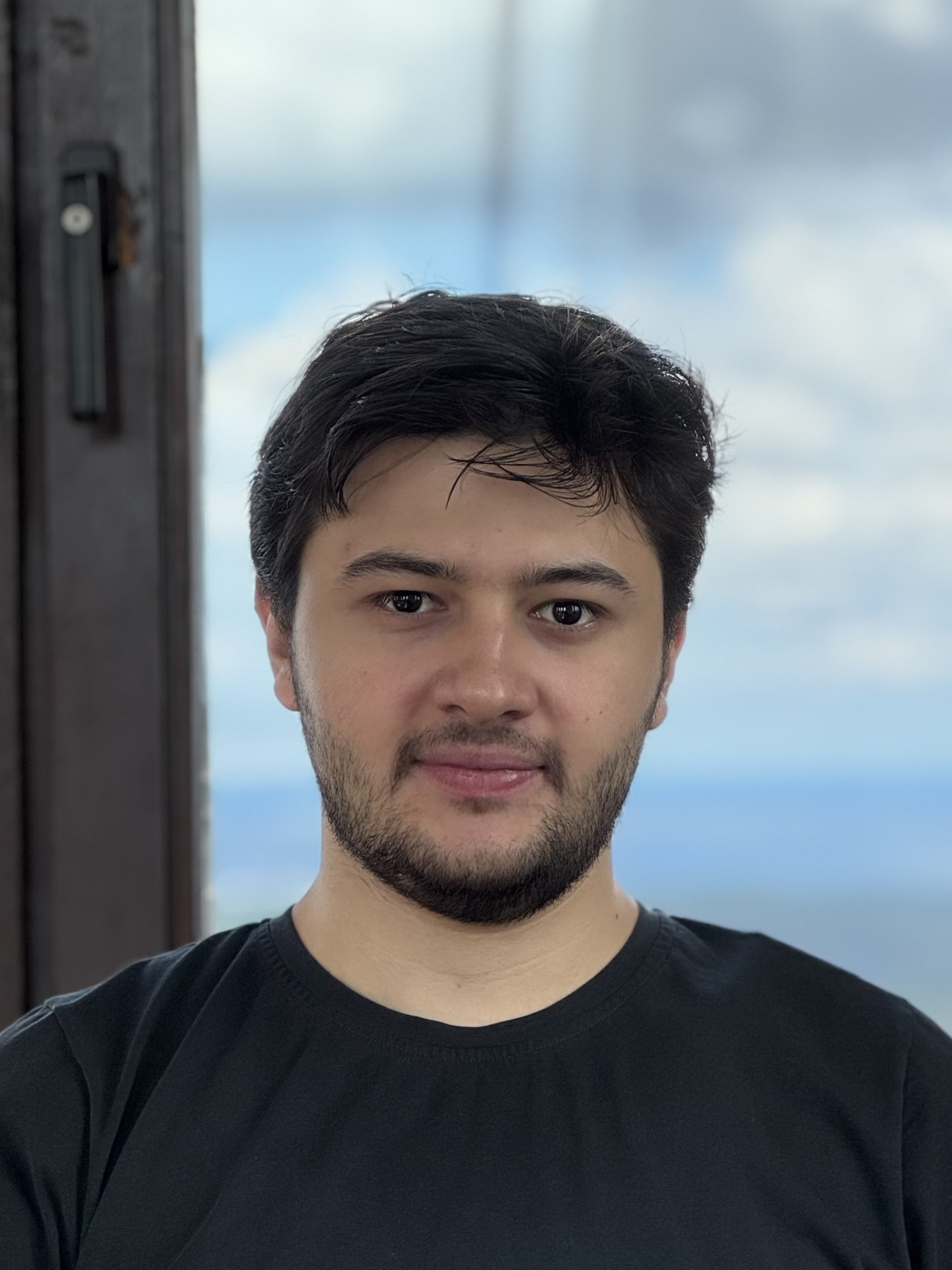}}]{Fatih Merdan}
completed his high school education at Kırıkkale Science High School. He received his B.Sc. degree in Electrical and Electronics Engineering from Middle East Technical University. He is currently pursuing his M.Sc. degree in Electrical and Electronics Engineering under the supervision of Prof. Akan at Koç University, Istanbul, Turkey.
\end{IEEEbiography}

\vspace{11pt}

\begin{IEEEbiography}[{\includegraphics[width=1in,height=1.25in,clip,keepaspectratio]{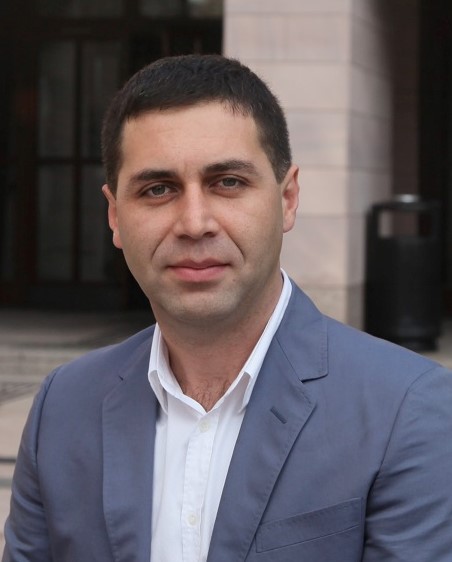}}]{Ozgur B. Akan}
\textbf{(Fellow, IEEE)} received the PhD
from the School of Electrical and Computer Engineering Georgia Institute of Technology Atlanta,
in 2004. He is currently the Head of Internet of
Everything (IoE) Group, with the Department of
Engineering, University of Cambridge, UK and the
Director of Centre for neXt-generation Communications (CXC), Koç University, Turkey. His research
interests include wireless, nano, and molecular communications and Internet of Everything.
\end{IEEEbiography}

\vfill

\end{document}